\documentclass[12pt]{article}
\usepackage{epsfig}
\usepackage{psfrag}
\usepackage{latexsym}
\def\beq{\begin{equation}}
\def\eeq{\end{equation}}
\def\bea{\begin{eqnarray}}
\def\eea{\end{eqnarray}}
\def\bet{\begin{tabular}}
\def\eet{\end{tabular}}
\def\bes{\begin{subequations}\bea}
\def\ees{\eea\end{subequations}}
\def\ol{\overline}

\def\e{\epsilon}

\def\be{\begin{equation}}
\def\ee{\end{equation}}
\def\bc{\begin{center}}
\def\ec{\end{center}}
\def\bea{\begin{eqnarray}}
\def\eea{\end{eqnarray}}
\def\dd{\displaystyle}
\def\nn{\nonumber}

\def\ov{\overline}

\catcode`@=11
\def\marginnote#1{}
\newcount\hour
\newcount\minute
\newtoks\amorpm
\hour=\time\divide\hour by60
\minute=\time{\multiply\hour by60 \global\advance\minute by-\hour}
\edef\standardtime{{\ifnum\hour<12 \global\amorpm={am}%
        \else\global\amorpm={pm}\advance\hour by-12 \fi
        \ifnum\hour=0 \hour=12 \fi
        \number\hour:\ifnum\minute<10 0\fi\number\minute\the\amorpm}}
\edef\militarytime{\number\hour:\ifnum\minute<10 0\fi\number\minute}
\def\draftlabel#1{{\@bsphack\if@filesw {\let\thepage\relax
   \xdef\@gtempa{\write\@auxout{\string
      \newlabel{#1}{{\@currentlabel}{\thepage}}}}}\@gtempa
   \if@nobreak \ifvmode\nobreak\fi\fi\fi\@esphack}
        \gdef\@eqnlabel{#1}}
\def\@eqnlabel{}
\def\@vacuum{}
\def\draftmarginnote#1{\marginpar{\raggedright\scriptsize\tt#1}}
\def\draft{\oddsidemargin 0.0truein
        \def\@oddfoot{\sl preliminary draft \hfil
        \rm\thepage\hfil\sl\today\quad\militarytime}
        \let\@evenfoot\@oddfoot \overfullrule 3pt
        \let\label=\draftlabel
        \let\marginnote=\draftmarginnote
   \def\@eqnnum{(\theequation)\rlap{\kern\marginparsep\tt\@eqnlabel}%
\global\let\@eqnlabel\@vacuum}  }
\catcode`@=12
\begin{document}
\begin{titlepage}
\vspace*{-1cm}
\phantom{hep-ph/0504165} 
\hfill{DFPD-05/TH/14}

\hfill{CERN-PH-TH/2005-067}

\vskip 0.5cm
\begin{center}
{\Large\bf Tri-Bimaximal Neutrino Mixing from Discrete Symmetry in Extra Dimensions}
\end{center}
\vskip 0.2  cm
\vskip 0.5  cm
\begin{center}
{\large Guido Altarelli}~\footnote{e-mail address: guido.altarelli@cern.ch}
\\
\vskip .1cm
CERN, Department of Physics, Theory Division
\\ 
CH-1211 Geneva 23, Switzerland
\\
\vskip .1cm
and
\\
Dipartimento di Fisica `E.~Amaldi', Universit\`a di Roma Tre
\\ 
INFN, Sezione di Roma Tre, I-00146 Rome, Italy
\\
\vskip .2cm
{\large Ferruccio Feruglio}~\footnote{e-mail address: feruglio@pd.infn.it}
\\
\vskip .1cm
Dipartimento di Fisica `G.~Galilei', Universit\`a di Padova 
\\ 
INFN, Sezione di Padova, Via Marzolo~8, I-35131 Padua, Italy
\\
\end{center}
\vskip 0.7cm
\begin{abstract}
\noindent
We discuss a particularly symmetric model of neutrino mixings where, with good accuracy, the atmospheric 
mixing angle $\theta_{23}$ is maximal, $\theta_{13}=0$ and the solar angle satisfies 
$\sin^2{\theta_{12}}=\frac{1}{3}$ (Harrison-Perkins-Scott (HPS) matrix). The discrete symmetry $A_4$ 
is a suitable symmetry group for the realization of this type of model. We construct a model where the 
HPS matrix is exactly obtained in a first approximation without imposing ad hoc relations among parameters.  
The crucial issue of the required VEV alignment in the scalar sector is discussed and we present a natural 
solution of this problem  based on a formulation with extra dimensions. We study the corrections from 
higher dimensionality operators allowed by the symmetries of the model and  discuss the conditions on the cut-off 
scales and the VEVs in order for these corrections to be completely under control. Finally, the observed hierarchy 
of charged lepton masses is obtained by assuming a larger flavour symmetry. 
We also show that, under general conditions, a maximal
$\theta_{23}$ can never arise from an exact flavour symmetry.
\end{abstract}
\end{titlepage}
\setcounter{footnote}{0}
\vskip2truecm
%
\section{Introduction}
By now there is convincing evidence for solar and atmospheric neutrino oscillations. The $\Delta m^2$ values and mixing angles are known with fair accuracy \cite{data}. For $\Delta m^2$ we have:  $\Delta m^2_{atm}\sim 2.5~10^{-3}$ eV$^2$ and  $\Delta m^2_{sol}\sim 8~10^{-5}$ eV$^2$. As for the mixing angles, two are large and one is small. The atmospheric angle $\theta_{23}$ is large, actually compatible with maximal but not necessarily so: at $3\sigma$: $0.31 \leq \sin^2{\theta_{23}}\leq 0.72$ with central value around  $0.5$. The solar angle $\theta_{12}$ is large, $\sin^2{\theta_{12}}\sim 0.3$, but certainly not maximal (by about 5-6 
$\sigma$ now \cite{lastSNO}). The third angle $\theta_{13}$ is strongly limited, mainly by the CHOOZ experiment, and has at present a $3\sigma$ upper limit given by about $\sin^2{\theta_{13}}\leq 0.08$. 

In spite of this experimental progress there are still many alternative routes in constructing models of neutrino masses. This variety is mostly due to the
considerable ambiguities that remain. First of all, it is essential to know whether
the LSND signal \cite{lsnd}, which has not been confirmed by KARMEN
\cite{karmen} and is currently being double-checked by MiniBoone
\cite{miniboone}, will be confirmed or will be excluded.  If LSND is right we probably
need at least four light neutrinos; if not we can do with only the three known ones, as we assume here in the following.  As
neutrino oscillations only determine mass squared differences a crucial missing input is the absolute scale of neutrino masses (within the existing limits from terrestrial experiments and cosmology \cite{Mainzetc},\cite{WMAPetc}).  Even for three neutrinos the pattern of the neutrino mass spectrum is still undetermined: it can be approximately degenerate, or of the inverse hierarchy type or normally hierarchical. Given for granted that neutrinos are Majorana particles, their masses can still arise either from the see-saw mechanism or from generic dimension-five non renormalizable operators. 

At a more direct level, we do not know how small the mixing angle $\theta_{13}$  is and how close to maximal is $\theta_{23}$. One can make a  distinction between "normal" and "special" models. For normal models  $\theta_{23}$ is not too close to maximal and $\theta_{13}$ is not too small, typically a small power of the self-suggesting order parameter $\sqrt{r}$, with $r=\Delta m_{sol}^2/\Delta m_{atm}^2 \sim 1/35$. Special models are those where some symmetry or dynamical feature assures in a natural way the near vanishing of $\theta_{13}$ and/or of $\theta_{23}- \pi/4$. Normal models are conceptually more economical and much simpler to construct. We expect that experiment will eventually find that  $\theta_{13}$ is not too small and that  $\theta_{23}$ is sizably not maximal. But if, on the contrary, either $\theta_{13}$ very small or  $\theta_{23}$ very close to maximal will emerge from experiment, then theory will need to cope with this fact. Thus it is  interesting to conceive and explore dynamical structures that could lead to special models in a natural way. 

We want to discuss here some particularly special models where both $\theta_{13}$ and $\theta_{23}- \pi/4$ exactly vanish
\footnote{More precisely,
they vanish in a suitable limit, with correction terms that can be made negligibly small.}. Then the neutrino mixing matrix $U_{fi}$ ($f=e$,$\mu$,$\tau$, $i=1,2,3$), in the basis of diagonal charged leptons, is given by, apart from sign convention redefinitions: 
\begin{equation}  
U_{fi}= 
\left(\matrix{ c_{12}&s_{12}&0 \cr  -s_{12}/\sqrt{2}&c_{12}/\sqrt{2}&-1/\sqrt{2}\cr
-s_{12}/\sqrt{2}&c_{12}/\sqrt{2}&1/\sqrt{2}     } \right) ~~~~~,
\label{ufi1}
\end{equation} 
where $c_{12}$ and $s_{12}$ stand for $\cos{\theta_{12}}$ and $\sin{\theta_{12}}$, respectively. It is much simpler to write natural models of this type with $s_{12}$ small and thus many such attempts are present in the early literature. More recently, given the experimental value of $\theta_{12}$, the more complicated case of $s_{12}$ large was also attacked, using non abelian symmetries, either continuous or discrete 
\cite{max1,max2,max3,max4,hps,max5,malast}. In many examples the invoked symmetries are particularly ad hoc and/or no sufficient attention is devoted to corrections from higher dimensional operators that can spoil the pattern arranged at tree level and to the highly non trivial vacuum alignment problems that arise if naturalness is required also at the level of vacuum expectation values (VEVs). 

An interesting special case of eq. (\ref{ufi1}) is obtained 
for  $s_{12}=1/\sqrt{3}$, i.e. the so-called tri-bimaximal or Harrison-Perkins-Scott mixing pattern  (HPS) 
\cite{hps}, with the entries in the second column all equal to $1/\sqrt{3}$ in absolute value:
\begin{equation}
U_{HPS}= \left(\matrix{
\dd\sqrt{\frac{2}{3}}&\dd\frac{1}{\sqrt 3}&0\cr
-\dd\frac{1}{\sqrt 6}&\dd\frac{1}{\sqrt 3}&-\dd\frac{1}{\sqrt 2}\cr
-\dd\frac{1}{\sqrt 6}&\dd\frac{1}{\sqrt 3}&\dd\frac{1}{\sqrt 2}}\right)~~~~~. 
\label{2}
\end{equation}
This matrix is a good approximation to present data
\footnote{In the HPS scheme $\tan^2{\theta_{12}}= 0.5$, to be compared with the latest experimental
determination \cite{lastSNO}: $\tan^2{\theta_{12}}= 0.45^{+0.09}_{-0.08}$.}. It would be interesting to find a natural and appealing scheme that leads to this matrix with good accuracy. In fact this is a most special model where not only $\theta_{13}$ and $\theta_{23}- \pi/4$ vanish but also $\theta_{12}$ assumes a particular value. Clearly, in a natural realization of this model, a very constraining and predictive dynamics must be underlying.  We think it is interesting to explore particular structures giving rise to this very special set of models in a natural way. In this case we have a maximum of "order" implying special values for all mixing angles: at the other extreme, anarchical models have been proposed \cite{anarchy}, where no structure at all is assumed in the lepton sector, so that, for example, $\theta_{13}$ and $\theta_{23}$ are predicted to be in no way special, except that there must be a smallest angle (probably near to the present bound) and a largest angle (expected sizably different from maximal). 

Interesting ideas on how to obtain the HPS mixing matrix have been discussed in refs \cite{hps}. 
The most attractive models 
are based on the discrete symmetry $A_4$, which appears as particularly suitable for the purpose, and were presented 
in ref. \cite{max3,malast}. In the present paper we start by discussing some general features of HPS models. 
We then present a new version of an $A_4$ model, with (moderate) normal hierarchy, and discuss in detail all aspects of naturalness in this model,  
also considering effects beyond tree level and the problem of vacuum alignment. There are a number of 
substantial improvements in our version with respect to Ma in ref. \cite{malast}. First, the HPS matrix is 
exactly obtained in a first approximation when higher dimensional operators are neglected, without imposing ad 
hoc relations among parameters (in ref. \cite{malast}. the equality of $b$ and $c$ is not guaranteed by the symmetry).
The observed hierarchy of charged lepton masses is obtained by assuming a larger flavour symmetry. 
The crucial issue of the required VEV alignment in the scalar sector is considered with special attention and we present a natural 
solution of this problem. We also keep the flavour scalar fields distinct from the normal Higgs bosons 
(a proliferation of Higgs doublets is disfavoured by coupling unification) and singlets under the Standard Model 
gauge group. Last not least, we study the corrections from higher dimensionality operators allowed by the 
symmetries of the model and  discuss the conditions on the cut-off scales and the VEVs in order for these 
corrections to be completely under control. 
%
%
\section{General Considerations}
The HPS mixing matrix implies that in a basis where charged lepton masses are 
diagonal the effective neutrino mass matrix is given by $m_{\nu}=U_{HPS}\rm{diag}(m_1,m_2,m_3)U_{HPS}^T$:
\begin{equation}
m_{\nu}=  \left[\frac{m_3}{2}\left(\matrix{
0&0&0\cr
0&1&-1\cr
0&-1&1}\right)+\frac{m_2}{3}\left(\matrix{
1&1&1\cr
1&1&1\cr
1&1&1}\right)+\frac{m_1}{6}\left(\matrix{
4&-2&-2\cr
-2&1&1\cr
-2&1&1}\right)\right]~~~~~. 
\label{1}
\end{equation}
The eigenvalues of $m_{\nu}$ are $m_1$, $m_2$, $m_3$ with eigenvectors $(-2,1,1)/\sqrt{6}$, $(1,1,1)/\sqrt{3}$ and $(0,1,-1)/\sqrt{2}$, respectively. In general, apart from phases, there are six parameters in a real symmetric matrix like $m_{\nu}$: here only three are left after the values of the three mixing angles have been fixed \`a la HPS. For a hierarchical spectrum $m_3>>m_2>>m_1$, $m_3^2 \sim \Delta m^2_{atm}$, $m_2^2/m_3^2 \sim \Delta m^2_{sol}/\Delta m^2_{atm}$ and $m_1$ could be negligible. But also degenerate masses and inverse hierarchy can be reproduced: for example, by taking $m_3= - m_2=m_1$  we have a degenerate model, while for $m_1= - m_2$ and $m_3=0$ an inverse hierarchy case (stability under renormalization group running strongly prefers opposite signs for the first and the second eigenvalue which are related to solar oscillations and have the smallest mass squared splitting). From the general expression of the eigenvectors one immediately sees that this mass matrix, independent of the values of $m_i$, leads to the HPS mixing matrix. It is a curiosity that the eigenvectors are the same as in the case of the Fritzsch-Xing (FX) matrix 
\cite{FX} but with the roles of the first and the third ones interchanged (so that for HPS $\theta_{23}$ is maximal while $\sin^2{2\theta_{12}}=8/9$, while for FX the two mixing angles keep the same values but are interchanged). 

If the atmospheric mixing angle is really maximal as in the HPS ansatz 
or close to maximal, 
it seems quite natural to interpret this as the effect of a 
flavour symmetry. It would be tempting to think of an approximate
flavour symmetry such that $\theta_{23}=\pi/4$ arises in the 
limit of exact symmetry, that is by neglecting all symmetry breaking 
effects.
Here we will show that this is not the case and that, under quite general
conditions, we can never obtain $\theta_{23}=\pi/4$ as a result of an 
{\em exact} flavour symmetry \footnote{For related observations see ref. 
\cite{nogo}.}. We assume that this symmetry is a 
meaningful symmetry, that is it is only broken by
small effects, in the real world.
In other words here we exclude symmetries that need
breaking terms of order one to describe the observed fermion masses
and mixing angles. Apart from that the symmetry can be of whatever
type, global or local, continuous or discrete.
Being interested in the limit of exact symmetry, we can 
neglect the sector giving rise to flavour symmetry breaking.
We assume that the fields on which such symmetry acts 
are the fields of the standard model, plus possibly the right-handed
neutrinos, so that our results will also cover the 
see-saw case.
Last, we assume canonical kinetic terms,
so that the symmetry acts on the fields of the standard model
through unitary transformations. 

Since the flavour symmetry is broken only by small
effects, the mass matrices for charged leptons and neutrinos 
can be written as:
\be
m_e=m_e^0+...~~~,~~~~~~~~~~m_\nu=m_\nu^0+...
\ee
where dots denote symmetry breaking effects and $m_e^0$
has rank less or equal than one.
Rank greater than one, as for instance when both
the tau and the muon have non-vanishing masses in the symmetry limit,
is clearly an unacceptable starting point, since the difference
between the two non-vanishing masses can only be explained by
large breaking effects, which we have excluded, or by a fine-tuning, 
which we wish to avoid.
If the rank of $m_e$ vanishes, than all mixing angles in the charged
lepton sector are undetermined in the symmetry limit and
$\theta_{23}$ is also completely undetermined.
Therefore we can focus on the case when $m_e^0$ has rank one.
If $m_e^0$ has rank one, then by a unitary transformations we can 
always go to a field basis where
\be
m_e^0=
\left(
\begin{array}{ccc}
0&0&0\cr
0&0&0\cr
0&0&m_\tau^0
\end{array}
\right)~~~.
\ee
As in the original basis, the action of the flavour symmetry on 
the new field basis is perfectly defined. If $U_\nu$ and $U_e$
are the unitary matrices that diagonalize $m_\nu^0$ and 
$m_e^{0\dagger} m_e^0$, it will be possible to adopt the parametrization 
\cite{king}
\be
U_\nu=K_\nu R_{23}(\theta^\nu_{23})P^\dagger
R_{13}(\theta^\nu_{13})P R_{12}(\theta^\nu_{12})~~~,
\ee
where $R_{ij}$ is the orthogonal matrix representing a rotation in the
$ij$ sector, $P={\rm diag}(1,1,\exp{i\delta})$ and
$K={\rm diag}(\exp{i\alpha_1},\exp{i\alpha_2},\exp{i\alpha_3})$.
Moreover:
\be
U_e=R_{12}(\theta^e_{12})~~~
\ee
where the angle $\theta^e_{12}$ is completely undetermined.
The physical mixing matrix is $U_{PMNS}=U_e^\dagger U_\nu$
and we find:
\be
\vert\tan\theta_{23}\vert=\vert\cos\theta^e_{12} \tan\theta^\nu_{23}
e^{i\alpha_2}+
\sin\theta^e_{12}\frac{\tan\theta^\nu_{13}}{\cos\theta^\nu_{23}}
e^{i(\delta+\alpha_1)}\vert~~~.
\label{tan23}
\ee
Therefore, in general, the atmospheric mixing angle is always
undetermined at the leading order. When small symmetry breaking
terms are added to $m_e^0$ and $m_\nu^0$, it is possible to
obtain $\theta_{23}=\pi/4$, provided these breaking
terms have suitable orientations in the flavour space.
If the breaking terms are produced by a spontaneous symmetry breaking
through the minimization of the potential energy of the theory,
in general two independent scalar sectors are needed.
One of them communicates the breaking to charged fermions
and the other one feeds the breaking to neutrinos.
In such a framework a maximal atmospheric mixing angle is always
the result of a special vacuum alignment.

In the literature
there are symmetries predicting $\theta_{23}$ large, not necessarily
maximal, in the limit of exact symmetry \cite{review}. For instance, this is 
produced by U(1) flavour symmetries, when the U(1) charges
of left-handed leptons and right-handed charged leptons are $(q_1,0,0)$ 
and $(p_1,p_2,0)$, respectively, with $q_1$ and $p_{1,2}$ all 
non-vanishing and different. In the symmetry limit, 
such an assignment implies ($m_e \sim \bar R L$, $m_{\nu} \sim L^TL$):
\be
m_e^{0\dagger} m_e^0=
\left(
\begin{array}{ccc}
0&0&0\cr
0&|\alpha|^2&\ol{\alpha}\beta\cr
0&\alpha\ol{\beta}&|\beta|^2
\end{array}
\right)~~~,
\ee
and:
\be
m_\nu^{0\dagger} m_\nu^0=
\left(
\begin{array}{ccc}
0&0&0\cr
0&|\alpha'|^2+|\gamma'|^2&\ol{\alpha'}\gamma'+\ol{\gamma'}\beta'\cr
0&\alpha'\ol{\gamma'}+\gamma'\ol{\beta'}&|\beta'|^2+|\gamma'|^2
\end{array}
\right)~~~,
\ee
with $\alpha$, $\beta$, $\alpha'$, $\beta'$ and $\gamma'$
independent parameters of the same order of magnitude.
If there is no conspiracy among these parameters,
the resulting $\theta_{23}$ mixing is generically
large. 

In conclusion, a large lepton mixing in the 23 sector
is possible as the result of an exact flavour symmetry.
But if we want to reproduce $\theta_{23}=\pi/4$ in some limit
of our theory, necessarily this limit cannot correspond
to an exact symmetry in flavour space. A maximal atmospheric
mixing angle can only originate from breaking effects
as a solution of a vacuum alignment problem.

%
%
\section{Basic Structure of the Model}

Our model is based on the discrete group $A_4$ following refs \cite{max3,malast}, where its structure and representations are described in detail. Here we simply recall that $A_4$ is the discrete symmetry group of the rotations that leave a tethraedron invariant, or the group of the even permutations of 4 objects. It has 12 elements and 4 inequivalent irreducible representations denoted 1, $1'$, $1''$ and 3  in terms of their respective dimensions. Introducing $\omega$, the cubic root of unity, $\omega=\exp{i\frac{2\pi}{3}}$, so that $1+\omega+\omega^2=0$, the three one-dimensional representations are obtained by dividing the 12 elements of $A_4$ in three classes, which are determined by the multiplication rule, and assigning to (class 1, class 2, class 3) a factor $(1,1,1)$ for 1, or $(1,\omega,\omega^2)$ for $1'$ or $(1,\omega^2,\omega)$ for $1''$. The product of two 3 gives $3 \times 3 = 1 + 1' + 1'' + 3 + 3$. Also $1' \times 1' = 1''$, $1' \times 1'' = 1$, $1'' \times 1'' = 1'$ etc.
For $3\sim (a_1,a_2,a_3)$, $3'\sim (b_1,b_2,b_3)$ the irreducible representations obtained from their product are:
\begin{equation}
1=a_1b_1+a_2b_2+a_3b_3
\end{equation}
\begin{equation}
1'=a_1b_1+\omega a_2b_2+\omega^2 a_3b_3
\end{equation}
\begin{equation}
1''=a_1b_1+\omega^2 a_2b_2+\omega a_3b_3
\end{equation}
\begin{equation}
3\sim (a_2b_3, a_3b_1, a_1b_2)
\end{equation}
\begin{equation}
3\sim (a_3b_2, a_1b_3, a_2b_1)
\end{equation}
Following ref. \cite{malast} we assigns leptons to the four inequivalent
representations of $A_4$: left-handed lepton doublets $l$ transform
as a triplet $3$, while the right-handed charged leptons $e^c$,
$\mu^c$ and $\tau^c$ transform as $1$, $1'$ and $1''$, respectively. 
The flavour symmetry is broken by two real triplets
$\varphi$ and $\varphi'$ and by a real singlet $\xi$. 
At variance with the choice made by \cite{malast}, these fields 
are gauge singlets.
Hence we only need two Higgs doublets $h_{u,d}$ (not three generations of
them as in ref. \cite{malast}), which we take invariant under $A_4$. 
We assume that some mechanism produces and maintains the hierarchy
$\langle h_{u,d}\rangle=v_{u,d}\ll \Lambda$ where $\Lambda$ is the 
cut-off scale of the theory
\footnote{This is the well known hierarchy 
problem that can be solved, for instance, by realizing a supersymmetric 
version of this model.}.
The Yukawa interactions in the lepton sector read:
\be
{\cal L}_Y=y_e e^c (\varphi l)+y_\mu \mu^c (\varphi l)''+
y_\tau \tau^c (\varphi l)'+ x_a\xi (ll)+x_d (\varphi' ll)+h.c.+...
\label{wl}
\ee
In our notation, $(3 3)$ transforms as $1$, 
$(3 3)'$ transforms as $1'$ and $(3 3)''$ transforms as $1''$.
Also, to keep our notation compact, we use a two-component notation
for the fermion fields and we set to 1 the Higgs fields
$h_{u,d}$ and the cut-off scale $\Lambda$. For instance 
$y_e e^c (\varphi l)$ stands for $y_e e^c (\varphi l) h_d/\Lambda$,
$x_a\xi (ll)$ stands for $x_a\xi (l h_u l h_u)/\Lambda^2$ and so on.
The Lagrangian  ${\cal L}_Y$ contains the lowest order operators
in an expansion in powers of $1/\Lambda$. Dots stand for higher
dimensional operators that will be discussed in section 6. 
Some terms allowed by the flavour symmetry, such as the terms 
obtained by the exchange $\varphi'\leftrightarrow \varphi$, 
or the term $(ll)$ are missing in ${\cal L}_Y$. 
Their absence is crucial and will be
motivated later on.

As we will demonstrate in section 5, the fields $\varphi'$,
$\varphi$ and $\xi$ develop a VEV along the directions:
\bea
\langle \varphi' \rangle&=&(v',0,0)\nn\\ 
\langle \varphi \rangle&=&(v,v,v)\nn\\
\langle \xi \rangle&=&u~~~. 
\label{align}
\eea 
Therefore, at the leading order of the $1/\Lambda$ expansion,
the mass matrices $m_l$ and $m_\nu$ for charged leptons and 
neutrinos are given by:
\be
m_l=v_d\frac{v}{\Lambda}\left(
\begin{array}{ccc}
y_e& y_e& y_e\\
y_\mu& y_\mu \omega& y_\mu \omega^2\\
y_\tau& y_\tau \omega^2& y_\tau \omega
\end{array}
\right)~~~,
\label{mch}
\ee
\be
m_\nu=\frac{v_u^2}{\Lambda}\left(
\begin{array}{ccc}
a& 0& 0\\
0& a& d\\
0& d& a
\end{array}
\right)~~~,
\label{mnu}
\ee
where
\be
a\equiv x_a\frac{u}{\Lambda}~~~,~~~~~~~d\equiv x_d\frac{v'}{\Lambda}~~~.
\label{ad}
\ee
Charged leptons are diagonalized by
\be
l\to \frac{1}{\sqrt{3}}\left(
\begin{array}{ccc}
1& 1& 1\\
1& \omega^2& \omega\\
1& \omega& \omega^2
\end{array}
\right)l~~~,
\label{change}
\ee
and charged fermion masses are given by:
\be
m_e=\sqrt{3} y_e v_d \frac{v}{\Lambda}~~~,~~~~~~~
m_\mu=\sqrt{3} y_\mu v_d \frac{v}{\Lambda}~~~,~~~~~~~
m_\tau=\sqrt{3} y_\tau v_d \frac{v}{\Lambda}~~~.
\label{chmasses}
\ee
We can easily obtain a natural hierarchy among $m_e$, $m_\mu$ and
$m_\tau$ by introducing an additional U(1)$_F$ flavour symmetry under
which only the right-handed lepton sector is charged.
We assign F-charges $0$, $2$ and $3\div 4$ to $\tau^c$, $\mu^c$ and
$\e^c$, respectively. By assuming that a flavon $\theta$, carrying
a negative unit of F, acquires a VEV 
$\langle \theta \rangle/\Lambda\equiv\lambda<1$, the Yukawa couplings
become field dependent quantities $y_{e,\mu,\tau}=y_{e,\mu,\tau}(\theta)$
and we have
\be
y_\tau\approx O(1)~~~,~~~~~~~y_\mu\approx O(\lambda^2)~~~,
~~~~~~~y_e\approx O(\lambda^{3\div 4})~~~.
\ee
In the flavour basis the neutrino mass matrix reads 
\footnote{Notice that a unitary change of basis like the one in eq. (\ref{change})
will in general change the relative phases of the eigenvalues of $m_\nu$.}:
\be
m_\nu=\frac{v_u^2}{\Lambda}\left(
\begin{array}{ccc}
a+2 d/3& -d/3& -d/3\\
-d/3& 2d/3& a-d/3\\
-d/3& a-d/3& 2 d/3
\end{array}
\right)~~~,
\label{mnu0}
\ee
and is diagonalized by the transformation:
\be
U^T m_\nu U =\frac{v_u^2}{\Lambda}{\tt diag}(a+d,a,-a+d)~~~,
\ee
with
\be
U=\left(
\begin{array}{ccc}
\sqrt{2/3}& 1/\sqrt{3}& 0\\
-1/\sqrt{6}& 1/\sqrt{3}& -1/\sqrt{2}\\
-1/\sqrt{6}& 1/\sqrt{3}& +1/\sqrt{2}
\end{array}
\right)~~~.
\ee
The leading order predictions are $\tan^2\theta_{23}=1$, 
$\tan^2\theta_{12}=0.5$ and $\theta_{13}=0$. The neutrino masses
are $m_1=a+d$, $m_2=a$ and $m_3=-a+d$, in units of $v_u^2/\Lambda$.
We can express $|a|$, $|d|$ in terms of 
$r\equiv \Delta m^2_{sol}/\Delta m^2_{atm}
\equiv (|m_2|^2-|m_1|^2)/|m_3|^2-|m_1|^2)$,
$\Delta m^2_{atm}\equiv|m_3|^2-|m_1|^2$ 
and $\cos\Delta$, $\Delta$ being the phase difference between
the complex numbers $a$ and $d$:
\bea
\sqrt{2}|a|\frac{v_u^2}{\Lambda}&=&
\frac{-\sqrt{\Delta m^2_{atm}}}{2 \cos\Delta\sqrt{1-2r}}\nn\\
\sqrt{2}|d|\frac{v_u^2}{\Lambda}&=&
\sqrt{1-2r}\sqrt{\Delta m^2_{atm}}~~~.
\label{tuning}
\eea
To satisfy these relations a moderate tuning is needed in our model.
Due to the absence of $(ll)$ in eq. (\ref{wl}) which we will motivate in the next section, $a$ and $d$ are of the same order in $1/\Lambda$, 
see eq. (\ref{ad}). Therefore we expect that $|a|$ and $|d|$ 
are close to each other and, to satisfy eqs. (\ref{tuning}),
$\cos\Delta$ should be negative and of order one. We obtain:
\bea
|m_1|^2&=&\left[-r+\frac{1}{8\cos^2\Delta(1-2r)}\right]
\Delta m^2_{atm}\nn\\
|m_2|^2&=&\frac{1}{8\cos^2\Delta(1-2r)}
\Delta m^2_{atm}\nn\\
|m_3|^2&=&\left[1-r+\frac{1}{8\cos^2\Delta(1-2r)}\right]\Delta m^2_{atm}
\label{lospe}
\eea
If $\cos\Delta=-1$, we have a neutrino spectrum close to hierarchical:
\be
|m_3|\approx 0.053~~{\rm eV}~~~,~~~~~~~
|m_1|\approx |m_2|\approx 0.017~~{\rm eV}~~~.
\ee 
In this case the sum of neutrino masses is about $0.087$ eV.
If $\cos\Delta$ is accidentally small, the neutrino spectrum becomes
degenerate. The value of $|m_{ee}|$, the parameter characterizing the 
violation of total lepton number in neutrinoless double beta decay,
is given by:
\be
|m_{ee}|^2=\left[-\frac{1+4 r}{9}+\frac{1}{8\cos^2\Delta(1-2r)}\right]
\Delta m^2_{atm}~~~.
\ee
For $\cos\Delta=-1$ we get $|m_{ee}|\approx 0.005$ eV, at the upper edge of
the range allowed for normal hierarchy, but unfortunately too small
to be detected in a near future.
Independently from the value of the unknown phase $\Delta$
we get the relation:
\be
|m_3|^2=|m_{ee}|^2+\frac{10}{9}\Delta m^2_{atm}\left(1-\frac{r}{2}\right)~~~,
\ee
which is a prediction of our model.

It is also important to get some constraint on the mass scales involved
in our construction. From eqs. (\ref{tuning}) and (\ref{ad}), 
by assuming $x_d\approx 1$ $v_u\approx 250$ GeV, we have
\be
\Lambda\approx 1.8\times 10^{15}~\left(\frac{v'}{\Lambda}\right)~~{\rm GeV}~~~.
\ee
Since, to have a meaningful expansion, we expect $v'\le\Lambda$,
we have the upper bound
\be
\Lambda< 1.8\times 10^{15}~~{\rm GeV}~~~.
\label{comax}
\ee
Beyond this energy scale, new physics should come into play.
The smaller the ratio $v'/\Lambda$, 
the smaller becomes the cut-off scale.
For instance, when $v'/\Lambda=0.03$, $\Lambda$ should be close to $10^{14}$
GeV. A complementary information comes from the charged lepton sector,
eq. (\ref{chmasses}). A lower bound on $v/\Lambda$ can be derived
from the requirement that the Yukawa coupling $y_\tau$ remains in
a perturbative regime. By asking $y_\tau v_d< 250$ GeV, we get
\be
\frac{v}{\Lambda}>0.004~~~.
\ee
Finally, by assuming that all the VEVs fall in approximately the same 
range, which will be shown in section 5, we obtain the range
\be
0.004<\frac{v'}{\Lambda}\approx \frac{v}{\Lambda}\approx 
\frac{u}{\Lambda}<1~~~,
\label{vsuL}
\ee that will be useful to estimate the effects of higher-dimensional
operators in section 6.
Correspondingly the cut-off scale will range between
about $10^{13}$ and $1.8\times 10^{15}$ GeV.
\vspace{0.5cm}
%
%
\section{Vacuum alignment}
In this section we investigate the problem of achieving the
vacuum alignment of eq. (\ref{align}). At the same time we
should prevent, at least at some level, the interchange
between the fields $\varphi$ and $\varphi'$ to produce
the desired mass matrices in the neutrino and charged lepton
sectors. As we will see, there are several difficulties
to naturally accomplish these requirements. 
By minimizing the scalar potential of the theory with
respect to $\varphi$ and $\varphi'$ we get six equations
that we would like to satisfy in terms of the two unknown
$v$ and $v'$. Even though we expect that, due to the symmetry $A_4$, 
the six minimum conditions are not necessarily independent,
such an expectation turns out to be wrong in the specific case, unless some additional
relation is enforced on the parameters of the scalar potential.
These additional relations are in general not natural.
For instance, even by imposing them at the tree level,
they are expected to be violated at the one-loop order.
Therefore, as we will now illustrate, the minimum conditions
cannot be all satisfied by our vacuum configuration.

As an example here we analyze the most general renormalizable
scalar potential invariant under $A_4$ and depending upon
the triplets $\varphi$ and $\varphi'$ of the Lagrangian
${\cal L}$ in eq. (\ref{wl}). The term $(ll)$ in
${\cal L}$ can be forbidden by an additional symmetry,
commuting with $A_4$. One possibility is just the total
lepton number $L$ or a discrete subgroup of it. Here we consider
a $Z_4$ symmetry under which $f^c$ transform
into $-i f^c$ $(f=e,\mu,\tau)$, $l$ into $i l$, $\varphi$ is invariant and $\varphi'$
changes sign. This symmetry also explains why $\varphi$ and $\varphi'$ cannot be interchanged.
The scalar potential $V$ contains bilinears $B_i$, trilinears $T_i$ 
and quartic terms $Q_i$, invariant under the group $A_4\times Z_4$. 
A choice of independent invariants is:
\bea
B_1&=&\varphi_1^2+\varphi_2^2+\varphi_3^2\nn\\
B_2&=&{\varphi'_1}^2+{\varphi'_2}^2+{\varphi'_3}^2\nn\\
T_1&=&\varphi_1\varphi_2\varphi_3\nn\\
T_2&=&\varphi_1\varphi'_2{\varphi'_3}+
\varphi_2\varphi'_3{\varphi'_1}+
\varphi_3\varphi'_1{\varphi'_2}\nn\\
Q_1&=&\varphi_1^2\varphi_2^2+\varphi_2^2\varphi_3^2
+\varphi_3^2\varphi_1^2\nn\\
Q_2&=&|\varphi_1^2+\omega^2\varphi_2^2+\omega\varphi_3^2|^2\nn\\
Q_3&=&{\varphi'_1}^2{\varphi'_2}^2+{\varphi'_2}^2{\varphi'_3}^2+
{\varphi'_3}^2{\varphi'_1}^2\nn\\
Q_4&=&|{\varphi'_1}^2+\omega^2 {\varphi'_2}^2
+\omega {\varphi'_3}^2|^2\nn\\
Q_5&=&\varphi_1\varphi_2\varphi'_1{\varphi'_2}+
\varphi_2\varphi_3\varphi'_2{\varphi'_3}+
\varphi_3\varphi_1\varphi'_3{\varphi'_1}\nn\\
Q_6&=&(\varphi_1^2+\varphi_2^2+\varphi_3^2)
({\varphi'_1}^2+{\varphi'_2}^2+{\varphi'_3}^2)
\nn\\
Q_7&=&(\varphi_1^2+\omega^2\varphi_2^2+\omega\varphi_3^2)
({\varphi'_1}^2+\omega {\varphi'_2}^2
+\omega^2 {\varphi'_3}^2)~~~,
\label{inv}
\eea
The scalar potential reads:
\bea
V&=&\frac{M_1^2}{2} B_1^2+\frac{M_2^2}{2} B_2^2
+\mu_1 T_1+\mu_2 T_2\nn\\
&+&c_1 Q_1+c_2 Q_2+c_3 Q_3+c_4 Q_4\nn\\
&+&c_5 Q_5 + c_6 Q_6
+(c_7 Q_7+c.c)~~~,
\eea
We start by analyzing the field configuration:
\bea
\langle \varphi \rangle&=&(v,v,v)\nn\\ 
\langle \varphi' \rangle&=&(v',0,0)
\label{align1}
\eea
The minimum conditions are:
\bea
\frac{\partial V}{\partial \varphi_1}&=&
M_1^2 v+\mu_1 v^2+4 c_1 v^3+2 c_6 v v'^2+2(c_7+{\overline c_7}) v v'^2=0\nn\\
\frac{\partial V}{\partial \varphi_2}&=&
M_1^2 v+\mu_1 v^2+4 c_1 v^3+2 c_6 v v'^2+2(\omega^2c_7+\omega{\overline c_7}) v v'^2=0\nn\\
\frac{\partial V}{\partial \varphi_3}&=&
M_1^2 v+\mu_1 v^2+4 c_1 v^3+2 c_6 v v'^2+2(\omega c_7+\omega^2{\overline c_7}) v v'^2=0\nn\\
\frac{\partial V}{\partial \varphi'_1}&=&
M_2^2 v'+4 c_4 v'^3+6 c_6 v^2 v'=0\nn\\
\frac{\partial V}{\partial \varphi'_2}&=&
\mu_2v v'+c_5 v^2 v'=0\nn\\
\frac{\partial V}{\partial \varphi'_3}&=&
\mu_2v v'+c_5 v^2 v'=0~~~.
\eea
The equations $\partial V/\partial \varphi_i=0$ are clearly 
incompatible unless $c_7=0$. Even by forcing $c_7$ to vanish,
we are left with three independent equations for the two unknown
$v$ and $v'$, which, for generic values of the coefficients,
admit only the trivial solution $v=v'=0$.
This negative results cannot be modified by adding to $V$ 
the terms depending on the singlet $\xi$.
Also by investigating the problem in a slightly more general
framework, with $\varphi$ real and $(\varphi',\xi)$ complex,
we reach the same conclusion. Although we have not a no-go
theorem, these examples show
the difficulty to obtain the desired alignment.

The difficulty illustrated above is not common to
all vacua. For instance the other possible alignment:
\bea
\langle \varphi \rangle&=&(v,v,v)\nn\\ 
\langle \varphi' \rangle&=&(v',v',v')
\label{align2}
\eea
leads to the minimum conditions:
\bea
\frac{\partial V}{\partial \varphi_i}&=&
M_1^2 v+\mu_1 v^2+\mu_2 v'^2+4 c_1 v^3+(2 c_5+6 c_6) v v'^2=0\nn\\
\frac{\partial V}{\partial \varphi'_i}&=&
M_2^2 v'+2 \mu_2 v v'+4 c_3 v'^3+(2 c_5+6 c_6) v^2 v'=0~~~.
\eea
In a non-vanishing portion of the parameter space, these
equations have non-trivial solution with non-vanishing $v$ and $v'$. 

It is possible to show that, by sufficiently restricting the form 
of the most general scalar potential invariant under $A_4$, the desired alignment
can be obtained. Restrictions that are unnatural in a generic model becomes
technically natural in a supersymmetric (SUSY) model. 
The well-known non-renormalization
properties of the superpotential allow to accept, at least from
a technical viewpoint, a restricted number of terms, compared to
what the $A_4$ symmetry would permit. Undesired terms of the
superpotential that are set to zero at the tree level
are not generated at any order in perturbation theory.
Indeed we have produced a SUSY example of this type, where the alignment problem
is solved and this example is discussed in detail
in the Appendix. However our real aim is to build a fully
natural model, where all the terms allowed by the symmetries
are present and where the only deviations from the symmetry limit
are provided by higher-dimensional operators, rather than by
small violations of ad-hoc imposed relations.
As we will now see, there exist a simple and economic solution
in the context of theories with one extra spatial dimension. 
%
%
\section{$A_4$ model in an extra dimension}
One of the problems we should overcome in the search for
the correct alignment is that of keeping neutrino and charged
lepton sectors separate, including the respective symmetry 
breaking sectors. Here we show that such a separation can be achieved
by means of an extra spatial dimension. The space-time is assumed
to be five-dimensional, the product of the four-dimensional
Minkowski space-time times an interval going from $y=0$ to $y=L$. 
At $y=0$ and $y=L$ the space-time has two four-dimensional boundaries,
which we will call branes. Our idea is that matter SU(2) singlets
such as $e^c,\mu^c,\tau^c$ are localized at $y=0$, while SU(2) doublets,
such as $l$ are localized at $y=L$ (see Fig.1). Neutrino masses
arise from local operators at $y=L$. Charged lepton
masses are produced by non-local effects involving both branes.
Later on we will see how such non-local effects can arise in this
theory. The simplest possibility is to introduce a bulk fermion,
depending on all space-time coordinates, that interacts with
$e^c,\mu^c,\tau^c$ at $y=0$ and with $l$ at $y=L$. The exchange of
such a fermion can provide the desired non-local coupling between
right-handed and left-handed ordinary fermions. Finally,
assuming that $\varphi$ and $(\varphi',\xi)$ are localized
respectively at $y=0$ and $y=L$, we obtain a natural separation
between the two sectors.
\begin{figure}[h!]
$$\hspace{-4mm}
\includegraphics[width=12.0 cm]{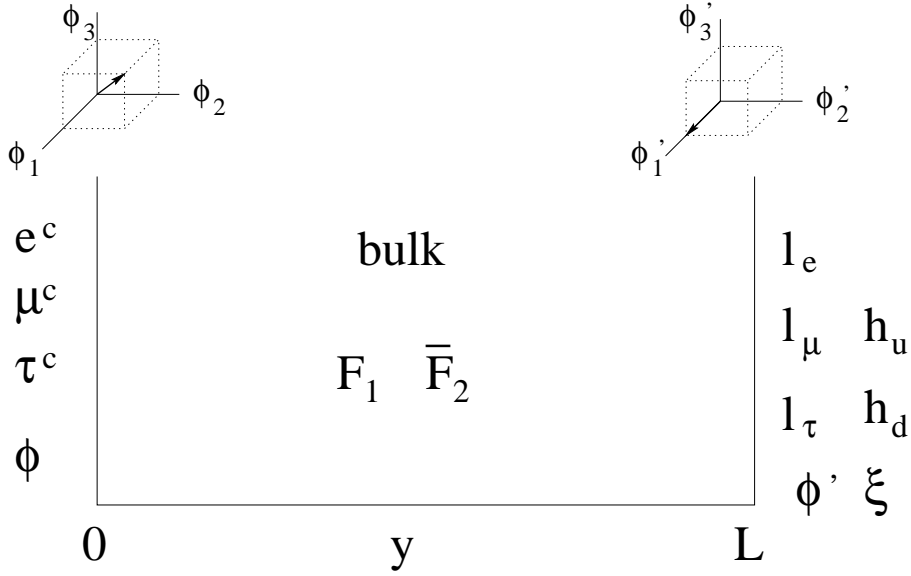}
$$
\caption[]
{Fifth dimension and localization of scalar and fermion fields.
The symmetry breaking sector includes the $A_4$ triplets $\varphi$
and $\varphi'$, localized at the opposite ends of the interval.
Their VEVs are dynamically aligned along the directions shown
at the top of the figure.}
\end{figure}

\subsection{Alignment in an extra dimension}
Such a separation also greatly simplify the vacuum alignment
problem. We can determine the minima of two
scalar potentials $V_0$ and $V_L$, depending only, respectively,
on $\varphi$ and $(\varphi',\xi)$. Indeed, as we shall see,
there are whole regions of the parameter space where 
$V_0(\varphi)$ and $V_L(\varphi',\xi)$
have the minima given in eq. (\ref{align}). Notice that in the present setup
dealing with a discrete symmetry such as $A_4$ provides a 
great advantage as far as the alignment problem 
is concerned. A continuous flavour symmetry such as, for instance,
SO(3) would need some extra structure to achieve the desired
alignment. Indeed the potential energy 
$\int d^4x [V_0(\varphi)+V_L(\varphi',\xi)]$ would be
invariant under a much bigger symmetry, SO(3)$_0\times$ SO(3)$_L$,
with the SO(3)$_0$ acting on $\varphi$ and leaving $(\varphi',\xi)$
invariant and vice-versa for SO(3)$_L$.
This symmetry would remove any alignment between the VEVs
of $\varphi$ and those of $(\varphi',\xi)$.
If, for instance, (\ref{align}) is minimum of the potential energy,
then any other configuration obtained by acting on (\ref{align})
with SO(3)$_0\times$ SO(3)$_L$ would also be a minimum
and the relative orientation between the two sets of VEVs would be
completely undetermined.
A discrete symmetry such as $A_4$ has not this problem, as we
will show now. 

Consider first the scalar potential $V_0(\varphi)$:
\be
V_0(\varphi)=\frac{M_1^2}{2}B_1^2+\mu_1 T_1+c_1 Q_1+c_2 Q_2~~~
\label{vzero}
\ee
where $B_1$, $T_1$, $Q_{1,2}$ are defined in eq. (\ref{inv}).
The minimum conditions at $\varphi=(v,v,v)$ are:
\be
\frac{\partial V_0}{\partial \varphi_i}=
v(M_1^2+\mu_1 v+4 c_1 v^2)=0~~~~~~~(i=1,2,3)~~~,
\ee
while the minimum condition at $\varphi=(v,0,0)$ is:
\be
\frac{\partial V_0}{\partial \varphi_1}=
v(M_1^2+4 c_2 v^2)=0~~~,
\ee
since in this case $(\partial V_0/\partial \varphi_{2,3})=0$
are automatically satisfied.
Both $\varphi=(v,v,v)$ and $\varphi=(v,0,0)$ can be local
minima of $V_0$, depending on the parameters. The constants 
$c_{1,2}$ should be positive, to have $V_0$ bounded from below.
We can look at the region where $|\mu_1|\ll |M_1|$.
When $c_1\gg c_2$ and $M_1^2<0$, the minimum at $\varphi=(v,0,0)$
is the absolute one, while for $c_2\gg c_1$ and $M_1^2<0$
$V_0$ is minimized by $\varphi=(v,v,v)$. 
Therefore we have a large portion of the parameter space
where the minimum is of the desired form: $\varphi=(v,v,v)$.
To be precise, in this region, there are four degenerate
minima: $\varphi=(v,v,v)$, $\varphi=(v,-v,-v)$ $\varphi=(-v,v,-v)$ 
$\varphi=(-v,-v,v)$, related by $A_4$ transformations.

Now we turn to $V_L(\varphi',\xi)$. As we did in section 4.1,
we assume both $\varphi'$ and $\xi$ real and odd under the action 
of a discrete $Z_4$ symmetry. The most general renormalizable
invariant potential is a combination of $B_2$, $Q_{3,4}$ in eq. 
(\ref{inv}) and the following invariants:
\bea
B_3&=&\xi^2\nn\\
Q_8&=&\xi^4\nn\\
Q_9&=&\xi\varphi'_1\varphi'_2\varphi'_3\nn\\
Q_{10}&=&\xi^2({\varphi'_1}^2+{\varphi'_2}^2+{\varphi'_3}^2)~~~.
\eea
We have:
\be
V_L(\varphi',\xi)=\frac{M_2^2}{2} B_2+\frac{M_3^2}{2} B_3+
c_3 Q_3+c_4 Q_4+c_8 Q_8+c_9 Q_9+c_{10} Q_{10}~~~.
\label{velle}
\ee
We search for a minimum at $\varphi'=(v',0,0)$ and $\xi=u$:
\bea
\frac{\partial V_L}{\partial \varphi'_1}&=&
v'(M_2^2+4 c_4 v'^2+ 2 c_{10} u^2)=0\nn\\
\frac{\partial V_L}{\partial \xi}&=&
u(M_3^2+4 c_8 u^2+2 c_{10} v'^2)=0~~~,
\eea
while $(\partial V_L/\partial \varphi'_{2,3})=0$
are always satisfied.
There is a region of the parameter space where the absolute
minimum is of this type. Taking into account the $A_4$
symmetry, in all this region we have six degenerate
minima: $\varphi'=(\pm v',0,0)$, $\varphi'=(0,\pm v',0)$ and 
$\varphi'=(0,0,\pm v')$.
Putting together the minima of $V_0(\varphi)$ and $V_L(\varphi',\xi)$
we have 24 degenerate minima of the potential energy, differing
for signs or ordering. It can be shown that these 24 minima
produce exactly the same mass pattern discussed in section 3.1,
up to field and parameter redefinitions. Therefore it is not restrictive
to choose one of them, for instance $\varphi=(v,v,v)$ and
$\varphi'=(v',0,0)$, to analyze the property of this model.

The observed hierarchy among lepton masses can be efficiently
described by an additional U(1)$_F$ flavour 
symmetry, under which only right-handed charged leptons are charged:
F$(e^c,\mu^c,\tau^c)=(4,2,0)$. To spontaneously break this symmetry
and to produce the desired hierarchy, we need a scalar field $\theta$,
carrying a negative unit of F and developing a VEV 
$\langle \theta \rangle/\Lambda\approx 0.22$. In our framework
$\theta$ is localized on the brane at $y=0$ and the scalar potential
$V_0$ of eq. (\ref{vzero}) is modified into:
\be
V_0\to V_0+M_4^2 B_4+c_{11} Q_{11}+c_{12} Q_{12}~~~,
\ee
where
\bea
B_4&=&|\theta|^2\nn\\
Q_{11}&=&|\theta|^4\nn\\
Q_{12}&=&|\theta|^2 (\varphi_1^2+\varphi_2^2+\varphi_3^2)~~~.
\eea
The minimum conditions at $\varphi=(v,v,v)$ and $|\theta|=t$ read:
\bea
\frac{\partial V_0}{\partial \varphi_i}&=&
v(M_1^2+\mu_1 v+4 c_1 v^2+2 c_{12} t^2)=0~~~~~~~(i=1,2,3)~~~,\nn\\
\frac{\partial V_0}{\partial |\theta|}&=&
2 t(M_4^2+2 c_{11}t^2+3 c_{12} v^2)=0~~~.
\eea
These conditions are satisfied by non-vanishing $(t,v)$ 
in a finite portion of the parameter space. 
Therefore the inclusion of an abelian flavour symmetry
is fully compatible with the mechanism for vacuum alignment 
discussed above.
\subsection{Lepton masses and mixing angles}
We now show how it is possible to take advantage of above
results to obtain the desired lepton masses. To this purpose
we introduce a bulk fermion field $F(x,y)=(F_1,\overline{F_2})$,
singlet under SU(2) with hypercharge $Y=-1$ and transforming
as a triplet of $A_4$. We also impose the discrete $Z_4$
symmetry introduced in section 4.1 under which $(f^c,l,F,\varphi,
\varphi',\xi)$ transform into 
$(-i f^c,i l,i F,\varphi,-\varphi',-\xi)$. The action is
\bea
S&=&\int d^4x dy \left\{
\left[
iF_1 \sigma^\mu \partial_\mu \ov{F}_1
+iF_2 \sigma^\mu \partial_\mu \ov{F}_2
+\frac{1}{2}(F_2\partial_y F_1-\partial_y F_2 F_1+h.c.)
\right]\right.\nn\\
&-&M(F_1F_2+\ov{F}_1\ov{F}_2)\nn\\
&+&V_0(\varphi)\delta(y)+V_L(\varphi',\xi)\delta(y-L)\nn\\
&+&\left[Y_e e^c (\varphi F_1)+Y_\mu \mu^c (\varphi F_1)''+
Y_\tau \tau^c (\varphi F_1)'+h.c.\right]\delta(y)\nn\\
&+&\left.\left[\frac{x_a}{\Lambda^2}\xi (ll)h_u h_u+
\frac{x_d}{\Lambda^2} (\varphi' ll)h_u h_u
+Y_L(F_2 l)h_d +h.c.\right]\delta(y-L)\right\}+...~~~,
\label{action5}
\eea
where the constants $Y$ have mass dimension -1/2.
The first two lines represent the five-dimensional kinetic and mass terms
of the bulk field $F$. The third line is the scalar potential 
and the remaining terms are the lowest order invariant operators
localized at the two branes. Dots stand for the kinetic terms
of $f^c,l,\varphi,\varphi',\xi$ and for higher-dimensional
operators, which will be classified in section 6.

The potential energy is given, at lowest order by:
\be
U=\int d^4x\left[V_0(\varphi)+V_L(\varphi',\xi)\right]~~~,
\label{penergy}
\ee
and, under the conditions discussed above, is minimized by
eqs. (\ref{align}).
It is clear that $\varphi$ and $(\varphi',\xi)$ are strictly 
separated only at lowest order. Indeed higher-dimensional brane
interactions like for instance $(\varphi\varphi F_1 F_2)/\Lambda^2$, 
$(\varphi'\varphi' F_1 F_2)/\Lambda^2$ are allowed. At the one-loop
level, the exchange of the bulk fermion $F$ will 
give rise to the structures $Q_{5,6,7}$ of eq. (\ref{inv})
and this will necessarily deform the vacuum (\ref{align}).
Here we will assume that such a deformation is sufficiently
small.  Indeed, as we shall see in section 6, the operators of the type 
$(\varphi\varphi\varphi'\varphi')$ arising from one-loop
F-exchange, are suppressed by $1/\Lambda^4 L^4$.

We now discuss the effects of the tree-level exchange of $F$.
To this purpose we consider the equations of motion for $(F_1,F_2)$:
\bea
i \sigma^\mu\partial_\mu\ov{F}_2+\partial_y F_1-M F_1&=&0\nn\\
i \sigma^\mu\partial_\mu\ov{F}_1-\partial_y F_2-M F_2&=&0
\eea
If $M$ is large and positive, we can prove that all the modes
contained in $(F_1,F_2)$ become heavy, at a scale greater than or comparable
to $1/L$, which we assume to be much higher than the electroweak scale.
If we are only interested in energies much lower than $1/L$,
we can solve the equations of motion in the static approximation,
by neglecting the four-dimensional kinetic term:
\bea
F_1(y)&=&F_1(L) e^{M(y-L)}\nn\\
F_2(y)&=&F_2(0) e^{-M y}~~~.
\label{sol}
\eea
These equations must be supplemented with appropriate boundary conditions,
which we can identify by varying the action $S$ with respect the fields
$(F_1,F_2)$. The boundary terms read
\bea
(\delta S)_{\rm boundary}&=&\int d^4x \left\{
\delta F_1(L)\left[\frac{1}{2} F_2(L)\right]\right.\nn\\
&+&\delta F_2(L)\left[-\frac{1}{2} F_1(L)+Y_L l h_d\right]\nn\\
&+&\delta F_1(0)\left[-\frac{1}{2} F_2(0)+Y_e e^c\varphi+
Y_\mu \mu^c \varphi_\mu+ Y_\tau \tau^c \varphi_\tau\right]\nn\\
&+&\left.\delta F_2(0)\left[\frac{1}{2} F_1(0)\right]\right\}~~~,
\label{bterms}
\eea
where $\varphi_\mu=(\varphi_1,\omega\varphi_2,\omega^2\varphi_3)$
and $\varphi_\tau=(\varphi_1,\omega^2\varphi_2,\omega\varphi_3)$.
We can chose as boundary conditions:
\bea
F_1(L)&=&2Y_L l h_d\nn\\
F_2(0)&=&2(Y_e e^c\varphi+
Y_\mu \mu^c \varphi_\mu+ Y_\tau \tau^c \varphi_\tau)~~~.
\label{bc}
\eea
Since $\delta F_1(L)=\delta F_2(0)=0$, we have $(\delta S)_{\rm boundary}=0$,
as desired. By substituting back eqs. (\ref{sol}) and (\ref{bc}) 
into the action $S$ we get
\bea
S=U&+&\int d^4x\left[\frac{y_e}{\Lambda} e^c (\varphi l)h_d
+\frac{y_\mu}{\Lambda} \mu^c (\varphi l)''h_d+
\frac{y_\tau}{\Lambda} \tau^c (\varphi l)'h_d\right.\nn\\
&+&\left.\frac{x_a}{\Lambda^2}\xi (ll)h_uh_u+
\frac{x_d}{\Lambda^2}(\varphi' ll)h_uh_u\right]+...~~~,
\label{slocal}
\eea
with 
\be
\frac{y_f}{\Lambda}=4 Y_L Y_f e^{- M L}~~~~~~~(f=e,\mu,\tau)~~~.
\label{yuka}
\ee
Therefore, in lowest order approximation we have reproduced the Lagrangian
${\cal L}_Y$ of eq. (\ref{wl}) and the discussion of section 3
applies.

We also recall that, to account for the observed hierarchy
of the charged lepton masses, we have included an additional
U(1) flavour symmetry. Therefore, in the present picture, the
quantities $Y_{e,\mu,\tau}$ stand for:
\be
Y_e=\tilde{Y}_e \left(\frac{\theta}{\Lambda}\right)^4~~~~~~~,~~~
Y_\mu=\tilde{Y}_\mu \left(\frac{\theta}{\Lambda}\right)^2~~~~~~~,~~~
Y_\tau=\tilde{Y}_\tau~~~,
\ee
where $\tilde{Y}_{e,\mu,\tau}$ are field-independent constants
having similar values. After spontaneous breaking of $U(1)$,
the Yukawa couplings $y_f$ possess the desired hierarchy.

%
\section{Higher-order corrections}
The results of the previous section hold to first approximation.
Higher-dimensional operators, suppressed by additional powers of 
the cut-off $\Lambda$, can be added to the leading terms in eqs.
(\ref{vzero},\ref{velle},\ref{penergy},\ref{slocal},\ref{yuka}).
Here we will classify these terms and analyze their
physical effects.  In particular we will show that these
corrections are completely under control in our model and that
they can be made negligibly small without any fine-tuning.
We can order higher-order operators
into three groups.
\subsection{Local corrections to $m_\nu$}
There are higher-order operators that are local in the 
five-dimensional theory and do not depend upon the heavy fermion sector
$(F_1,\ol{F}_2)$. As we have seen, at leading order, the neutrino 
mass matrix $m_\nu$ arises entirely from operators of this type
that are localized at $y=L$. On this brane we only have
scalar fields $(\xi,\varphi')$, odd under $Z_4$. Therefore
higher-dimensional operators modifying $m_\nu$ and localized 
at $y=L$ are down by two powers of the cut-off, compared to the 
leading ones. After $A_4$ breaking, the only two operators
that cannot be absorbed by a redefinition of the parameters
$x_{a,d}$ are:
\be
\begin{array}{c}
\dd\frac{x_b}{\Lambda^4}\xi(\varphi'\varphi')'(ll)''h_uh_u\\
\\
\dd\frac{x_c}{\Lambda^4}\xi(\varphi'\varphi')''(ll)'h_uh_u
\end{array}~~~.
\label{hbc}
\ee
After adding these operators localized at $y=L$ to the 
five-dimensional action of eq. (\ref{action5}), we get a
neutrino mass matrix
\be
m_\nu=\frac{v_u^2}{\Lambda}
\left(
\begin{array}{ccc}
a+b+c&0&0\\
0&a+\omega b+\omega^2 c&d\\
0&d&a+\omega^2 b+\omega c
\end{array}
\right)~~~,
\ee
where
\be
b\equiv x_b\frac{uv'^2}{\Lambda^3}~~~,~~~~~~~
c\equiv x_c\frac{uv'^2}{\Lambda^3}~~~,
\ee
to be compared with $a$ and $d$ of eq. (\ref{ad}).
\subsection{Corrections from tree-level $F$-exchange}
Another set of higher-dimensional operators arise from
the exchange of the heavy fermion $(F_1,\ol{F}_2)$ in the static limit
and in the tree-level approximation. To classify them,
we should list all operators localized at the two branes 
that are linear in the bulk fermion $(F_1,\ol{F}_2)$. At $y=0$ such
operators have the generic structure
\be
Y_f f^c \varphi F_1~~~,
~~~~~~~\frac{Y^{(1)}_f}{\Lambda} f^c \varphi^2 F_1~~~,
~~~~~~~\frac{Y^{(2)}_f}{\Lambda^2} f^c \varphi^3 F_1~~~...
~~~~~~~(f=e,\mu,\tau)
\ee
After spontaneous $A_4$ breaking, the effect of these operators
can be absorbed by redefining the coupling constants
$Y_f$, $(f=e,\mu,\tau)$, at least up to order $\varphi^3$. Thus
the leading interactions between $f^c$ and $F_1$
\be
\left[Y_e e^c (\varphi F_1)+Y_\mu \mu^c (\varphi F_1)''+
Y_\tau \tau^c (\varphi F_1)'+h.c.\right]\delta(y)
\ee
are unchanged up to relative order $1/\Lambda^2$.
We are left with the couplings of $F_2$ at the brane $y=L$.
Neglecting all operators that, after $A_4$ breaking, only
lead to a renormalization of the parameter $Y_L$, we find
four new terms:
\be
\begin{array}{c}
\dd\frac{Z_1}{\Lambda^2}(\varphi'\varphi')'(F_2 l)''h_d\\
\\
\dd\frac{Z_2}{\Lambda^2}(\varphi'\varphi')''(F_2 l)'h_d\\
\\
\dd\frac{Z_3}{\Lambda^2}\xi
\left[\varphi'_1 (F_2)_2 l_3+
\varphi'_2 (F_2)_3 l_1+
\varphi'_3 (F_2)_1 l_2\right]h_d\\
\\
\dd\frac{Z_4}{\Lambda^2}\xi
\left[\varphi'_1 (F_2)_3 l_2+
\varphi'_2 (F_2)_1 l_3+
\varphi'_3 (F_2)_2 l_1\right]h_d
\end{array}
\label{zop}
\ee
After the breaking of $A_4$, the leading order interaction
of $F_2$ at $y=L$ is modified by the operators (\ref{zop}) to
\be
\left[Y_L(F_2 \tilde{l})h_d +h.c.\right]\delta(y-L)
\ee
where
\be
\left(
\begin{array}{c}
\tilde{l}_1\\
\tilde{l}_2\\
\tilde{l}_3
\end{array}
\right)
=
\left(
\begin{array}{ccc}
1+z_1+z_2&0&0\\
0&1+\omega z_1+\omega^2 z_2&z_4\\
0&z_3&1+\omega^2 z_1+\omega z_2
\end{array}
\right)
\left(
\begin{array}{c}
{l}_1\\
{l}_2\\
{l}_3
\end{array}
\right)
\ee
\be
z_{1,2}\equiv \frac{Z_{1,2}}{Y_L}\frac{v'^2}{\Lambda^2}~~~,~~~~~~~
z_{3,4}\equiv \frac{Z_{3,4}}{Y_L}\frac{u v'}{\Lambda^2}~~~.
\ee
After integrating out the heavy modes in $(F_1,\ol{F}_2)$
in the limit of vanishing external momenta for the light modes,
we obtain the effective four-dimensional Lagrangian
\be
\frac{y_e}{\Lambda} e^c (\varphi \tilde{l})h_d
+\frac{y_\mu}{\Lambda} \mu^c (\varphi \tilde{l})''h_d+
\frac{y_\tau}{\Lambda} \tau^c (\varphi \tilde{l})'h_d~~~,
\ee
\be
\frac{y_f}{\Lambda}=4 Y_L Y_f e^{- M L}~~~~~~~(f=e,\mu,\tau)~~~.
\ee
The mass matrix for the charged leptons becomes
\be
m_l=v_d\frac{v}{\Lambda}\left(
\begin{array}{ccc}
y_e(1+z_1+z_2)& y_e(1+\omega z_1+\omega^2 z_2+z_3)& 
y_e(1+\omega^2 z_1+\omega z_2+z_4)\\
y_\mu(1+z_1+z_2)& y_\mu \omega(1+\omega z_1+\omega^2 z_2+\omega z_3) & 
y_\mu \omega^2(1+\omega^2 z_1+\omega z_2+\omega^2 z_4)\\
y_\tau(1+z_1+z_2)& y_\tau \omega^2(1+\omega z_1+\omega^2 z_2+\omega^2 z_3)& 
y_\tau \omega(1+\omega^2 z_1+\omega z_2+\omega z_4)
\end{array}
\right)~~~.
\ee
\subsection{Effects on masses and mixing angles}
To first order in the small parameters $b$, $c$ and $z_i$, the 
neutrino masses are modified into:
\bea
m_1&=&\left(a+d-\frac{1}{2}(b+c)\right)\frac{v_u^2}{\Lambda}\nn\\
m_2&=&\left(a+b+c\right)\frac{v_u^2}{\Lambda}\nn\\
m_3&=&\left(-a+d+\frac{1}{2}(b+c)\right)\frac{v_u^2}{\Lambda}
\eea
and the charged lepton masses are changed into
\bea
m_e&=&\sqrt{3} y_e (1+\frac{z_3}{3}+\frac{z_4}{3})v_d\frac{v}{\Lambda}\nn\\
m_\mu&=&\sqrt{3} y_\mu (1+\omega\frac{z_3}{3}+\omega^2\frac{z_4}{3})
v_d\frac{v}{\Lambda}\nn\\
m_\tau&=&\sqrt{3} y_\tau(1+\omega^2\frac{z_3}{3}+\omega\frac{z_4}{3})
v_d\frac{v}{\Lambda}
\eea
To the same order, but neglecting terms like $z_i~ y_{e,\mu}/y_\tau$,
we get:
\bea
|U_{e3}|&=&\left\vert\frac{(\ol{b}-\ol{c})(d-a)+(b-c)(\ol{d}+\ol{a})}
{2\sqrt{2}(a\ol{d}+\ol{a}d)}+
\frac{1}{\sqrt{2}}(-\ol{z}_1+\ol{z}_2+\frac{i}{\sqrt{3}}{z}_3-
\frac{i}{\sqrt{3}}{z}_4)\right\vert\nn\\
|\tan^2\theta_{23}|&=&1+\frac{(\ol{b}-\ol{c})d+(b-c)\ol{d}}{(a\ol{d}+\ol{a}d)}
+2\left[z_2+\ol{z}_2+\frac{1}{3}(z_3+\ol{z}_3+z_4+\ol{z}_4)\right]\nn\\
|\tan^2\theta_{12}|&=&\frac{1}{2}\left[1+\frac{3}{2}\left(
-z_1-\ol{z}_1-z_2-\ol{z}_2+\frac{z_3+\ol{z}_3}{3}+\frac{z_4+\ol{z}_4}{3}
\right)\right]
\label{corr}
\eea
These relations explicitly show that the corrections
induced by the higher-dimensional operators are of order $uv'/\Lambda^2$ or
$v'^2/\Lambda^2$. From our estimate in eq. (\ref{vsuL}) we see that
these parameters can be as small as $2~10^{-5}$. If the cut-off $\Lambda$
is one order of magnitude larger that the VEVs of the model,
the resulting corrections are at the level of one percent, already
beyond any planned experimental test. If on the contrary, the
VEVs are anomalously close to the cut-off $\Lambda$, then eqs. (\ref{corr})
show that deviations roughly of the same size are expected in 
$U_{e3}$, $\tan^2\theta_{23}$ and $\tan^2\theta_{12}$.
How much close to $\Lambda$ can the VEVs be? We expect that the 
subleading corrections do not spoil the leading order form of
the neutrino mass spectrum, eq. (\ref{lospe}). This implies
that $v'^2/\Lambda^2\ll r$, so that $r$ sets the natural upper bound
to the expected deviations from the leading order results.
\subsection{Corrections from one-loop $F$-exchange}
Further corrections to lepton mass matrices and to the scalar
potential can arise from one-loop exchange of $(F_1,\ol{F}_2)$ 
in the static limit. Consider for instance the following
operators localized at $y=0$ and at $y=L$:
\be
\frac{1}{\Lambda}\varphi F_1 F_2\delta(y)~~~,~~~~~~~
\frac{1}{\Lambda^6}\xi l l F_1 F_2 h_u h_u\delta(y-L)~~~.
\ee
By integrating out, at one-loop order, the heavy modes contained
in $(F_1,\ov{F}_2)$ we get:
\be
\frac{1}{\Lambda^7}\xi\varphi l l h_u h_u\int d^4 k 
\Delta_F(k,0,L)\Delta_F(k,L,0)~~~,
\ee
where $k$ is the four-momentum running in the loop and
$\Delta_F(k,y,y')$ is the adimensional propagator of $(F_1,\ov{F}_2)$
in a mixed momentum-space representation. Since the loop integral
is convergent, we get
\be
\frac{1}{\Lambda^3}\frac{f(ML)}{\Lambda^4 L^4}\xi\varphi l l h_u h_u ~~~,
\ee
where $f(ML)$ is a function of the adimensional combination $M L$.
Thus the resulting local operator is suppressed by four additional powers 
of the cut-off scale. This behavior is quite generic and similar
suppressions are found for other operators originating from one-loop
exchange of $(F_1,\ov{F}_2)$. 

The corrections that modify the scalar potential discussed 
in the previous section are of this type. 
As an example, consider the localized interactions:
\be
\frac{1}{\Lambda^2}\varphi\varphi F_1 F_2\delta(y)~~~,~~~~~~~
\frac{1}{\Lambda^2}\varphi'\varphi' F_1 F_2\delta(y-L)~~~.
\ee
Also in this case, after integrating over $(F_1,\ov{F}_2)$ in the
limit of vanishing external momenta, we get:
\be
\frac{f(ML)}{\Lambda^4 L^4}\varphi\varphi\varphi'\varphi'~~~.
\ee
Due to their large suppression, these corrections are negligible
compared to those discussed above.
\vskip 0.2 cm
\section{Conclusion}
There are by now several theoretical mechanisms that can
qualitatively explain the observed large lepton mixing angles \cite{review}.
They are sufficiently flexible to quantitatively accommodate
the measured parameters. They are also
compatible with our ideas on quarks masses and mixing angles
so that they can be nicely embedded into a unified picture
of fermion properties, such as, for instance, a grand unified 
theory. Many of these mechanisms predict a generically large
atmospheric mixing angle and a generically small
$\theta_{13}$ angle, without favouring any specific value
for these parameters. The best values of global fits
are currently very close to $\theta_{23}=\pi/4$ and
$\theta_{13}=0$, but the experimental errors
still allow for large deviations from these remarkable values.
Indeed, according to many of the above mentioned mechanisms,
deviations from $\theta_{23}=\pi/4$ and
$\theta_{13}=0$ are expected at the observable level. 
It may take a long time before
such deviations can be actually observed. A sensitivity on $\theta_{13}$ 
around $0.05$ is foreseen in about ten years from now, with
the full exploitation of high-intensity neutrino beams.
A reduction by a factor of two of the present error on $\theta_{23}$
will also require special neutrino beams and a similar time scale.

It might happen that after all this experimental effort, 
$(\theta_{23}-\pi/4)$ and
$\theta_{13}$ still remain close to zero, within errors. 
At this point it would be legitimate to suspect that
such special values are produced by a highly symmetric
flavour dynamics. 
Given the already good experimental precision on $\theta_{12}$,
the so-called
Harrison-Perkins-Scott mixing scheme, where $\theta_{23}=\pi/4$,
$\theta_{13}=0$ and $\sin^2\theta_{12}=1/3$, would fit very well
the data. In this paper we have proposed a model that reproduces
accurately the HPS mixing pattern.
We started by discussing whether such a pattern 
can be obtained from an exact flavour symmetry. 
We showed that, under general conditions, 
an exactly maximal atmospheric mixing angle
cannot arise from an exact flavour symmetry.
The flavour symmetry should 
be necessarily broken and a maximal $\theta_{23}$ is the result of a 
special alignment between the breaking effects in the neutrino
sector and those occurring in the charged lepton sector.
If the flavour symmetry is spontaneously broken,
this corresponds to a non-trivial vacuum alignment.
Our model gives rise to the HPS mixing scheme
in the context of a spontaneously broken $A_4$
flavour symmetry, $A_4$ being the discrete subgroup
of SO(3) leaving a tetrahedron invariant.

At leading order, that is by neglecting symmetric operators
of higher dimension, neutrino masses only depend on two 
complex Yukawa coupling constants. Due to the unknown phase
difference between these two constants, we cannot determine
the absolute scale of neutrino masses. We expect
that the neutrino spectrum is of the normal hierarchy type but 
not too far from degenerate. At leading order the model predicts
$|m_3|^2=|m_{ee}|^2+10/9\Delta m^2_{atm}(1-r/2)$.
A remarkable feature of our model is that at the leading order 
the lepton mixing angles are completely independent from these two parameters, 
so that the HPS mixing pattern is always obtained.
The lepton mixing depends entirely on the relative
alignment between the VEVs giving masses to the neutrino
sector and those giving masses to the lepton sector.
We discuss in detail the problem of vacuum alignment.
To avoid the proliferation of Higgs doublets, the scalar fields
breaking $A_4$ are gauge singlets in our model.
We propose an unconventional solution to the vacuum alignment problem,
where an extra dimension described by a spatial interval plays
an important role.
Two scalar sectors live at the opposite ends of the interval
and their respective scalar potentials are minimized by the 
desired field configurations, for natural values of the implied parameters.
Such a mechanism only works in the case of discrete symmetries,
since in the continuous case the large symmetry of the total
potential energy would make the relative
orientations of the two scalar sectors undetermined.
We have also extensively discussed how this lowest order picture is modified by
the introduction of higher dimensional operators.
The induced corrections are parametrically small, of second
order in the expansion parameter $VEV/\Lambda$, $\Lambda$ being the
cut-off of the theory, and they can be made
numerically negligible.
Last but not least, the hierarchy of the charged lepton masses
can be reproduced by the usual Froggatt-Nielsen mechanism
within the context of an abelian flavour symmetry, which
turns out to be fully compatible with the present scheme.

We believe that, from a purely technical point of view,  we have fulfilled our goal to realize a completely natural construction of the HPS mixing scheme. But to construct our model we had to introduce a number of special dynamical tricks (like a peculiar set of discrete symmetries in extra dimensions). Apparently this is  the price to pay for a ``special'' model where all mixing angles are fixed to particular values. Perhaps this exercise can be taken as a hint that it is more plausible to expect that, in the end, experiment will select a ``normal'' model with $\theta_{13}$ not too small and $\theta_{23}$ not too close to maximal.

\vskip 0.2 cm
\section*{Acknowledgment}
We thank Zurab Berezhiani, Isabella Masina and Luigi Pilo for useful discussions. F.F. thanks the CERN Theory Division
for hospitality in summer 2004, when this project started. This project is partially
supported by the European Program MRTN-CT-2004-503369.
\vskip 0.2 cm
\vfill
\newpage
\section*{Appendix}
%
%
Here we discuss a SUSY solution to the vacuum alignment problem.
In a supersymmetric context, the right-hand side of eq. (\ref{wl})
should be interpreted as the superpotential $w_l$ of the theory,
in the lepton sector.
A key observation is that this superpotential 
is invariant not only with respect to the gauge symmetry 
SU(2)$\times$ U(1) and the flavour symmetry U(1)$_F\times A_4$,
but also under a discrete $Z_3$ symmetry and a continuous U(1)$_R$ 
symmetry under which the fields 
transform as shown in table 1.
\\[0.2cm]
\begin{center}
\begin{tabular}{|c||c|c|c|c||c|c|c|c||c|c|c|}
\hline
{\tt Field}& l & $e^c$ & $\mu^c$ & $\tau^c$ & $h_{u,d}$ & 
$\varphi$ & $\varphi'$ & $\xi$ & $\varphi_0$ & $\varphi'_0$ & $\xi_0$\\
\hline
$A_4$ & $3$ & $1$ & $1'$ & $1''$ & $1$ & 
$3$ & $3$ & $1$ & $3$ & $3$ & $1$\\
\hline
$Z_3$ & $\omega$ & $\omega^2$ & $\omega^2$ & $\omega^2$ & $1$ &
$1$ & $\omega$ & $\omega$ & $1$ & $\omega$ & $\omega$\\
\hline
$U(1)_R$ & $1$ & $1$ & $1$ & $1$ & $0$ & 
$0$ & $0$ & $0$ & $2$ & $2$ & $2$\\
\hline
\end{tabular}
\end{center}
\vspace{0.2cm}
We see that the $Z_3$ symmetry explains the absence of the term $(ll)$
in $w_l$: such a term transforms as $\omega^2$ under $Z_3$ and
need to be compensated by the field $\xi$ in our construction.
At the same time $Z_3$ does not allow the interchange between
$\varphi'$ and $\varphi$, which transform differently under $Z_3$. 
Charged leptons and neutrinos acquire
masses from two independent sets of fields. 
If the two sets of fields develop VEVs according to the 
alignment described in eq. (\ref{align}), then the desired
mass matrices follow.

Finally, there is a continuous $U(1)_R$
symmetry that contains the usual $R$-parity as a subgroup.
Suitably extended to the quark sector, this symmetry forbids
the unwanted dimension two and three terms in the superpotential
that violate baryon and lepton number at the renormalizable level. 
The $U(1)_R$ symmetry allows us to classify
fields into three sectors. There are ``matter fields'' such as the 
leptons $l$, $e^c$, $\mu^c$ and $\tau^c$, which occur in the 
superpotential through bilinear combinations. There is a 
``symmetry breaking sector'' including the higgs doublets
$h_{u,d}$ and the flavons $\varphi'$, $\varphi$ and $\xi$.
As we will see these fields acquire non-vanishing vacuum expectation
values (VEVs) and break the symmetries of the model.
Finally, there are ``driving fields'' such as $\varphi'_0$, $\varphi_0$
and $\xi_0$ that allows to build a 
non-trivial scalar potential in the symmetry breaking sector. 
Since driving fields have R-charge equal to two, the superpotential
is linear in these fields.

The full superpotential of the model is
\be
w=w_l+w_d
\ee 
where, at leading order in a $1/\Lambda$ expansion, $w_l$ is given
by the right-hand side of eq. (\ref{wl}) and the ``driving'' term 
$w_d$ reads:
\be
w_d=M (\varphi_0 \varphi)+ g (\varphi_0 \varphi\varphi)+
g_1 (\varphi'_0 \varphi'\varphi')+
g_2 \xi (\varphi'_0\varphi')+
g_3 \xi_0 (\varphi'\varphi')+
g_4 \xi_0 \xi^2~~~.
\ee
We notice that at the leading order there are no terms involving
the Higgs fields $h_{u,d}$. We assume that the electroweak symmetry
is broken by some mechanism, such as radiative effects when supersymmetry
(SUSY) is broken. It is interesting that at the leading order
the electroweak scale does not mix with the potentially large scales
$u$, $v$ and $v'$. The scalar potential is given by:
\be
V=\sum_i\left\vert\frac{\partial w}{\partial \phi_i}\right\vert^2
+m_i^2 \vert \phi_i\vert^2+...
\ee
where $\phi_i$ denote collectively all the scalar fields of the 
theory, $m_i^2$ are soft masses and dots stand for D-terms for the 
fields charged under the gauge group and possible additional
soft breaking terms. Since $m_i$ are expected to be much smaller
than the mass scales involved in $w_d$, it makes sense to
minimize $V$ in the supersymmetric limit and to account for soft 
breaking effects subsequently. From the driving sector we have:
\bea
\frac{\partial w}{\partial \varphi_{01}}&=&M\varphi_1
+g\varphi_2\varphi_3=0\nn\\
\frac{\partial w}{\partial \varphi_{02}}&=&M\varphi_2
+g\varphi_3\varphi_1=0\nn\\
\frac{\partial w}{\partial \varphi_{03}}&=&M\varphi_3
+g\varphi_1\varphi_2=0\nn\\
\frac{\partial w}{\partial \varphi'_{01}}&=&
g_1\varphi'_2\varphi'_3+g_2\xi \varphi'_1=0\nn\\
\frac{\partial w}{\partial \varphi'_{02}}&=&
g_1\varphi'_3\varphi'_1+g_2\xi \varphi'_2=0\nn\\
\frac{\partial w}{\partial \varphi'_{03}}&=&
g_1\varphi'_1\varphi'_2+g_2\xi \varphi'_3=0\nn\\
\frac{\partial w}{\partial \xi_0}&=&
g_3(\varphi'\varphi')+g_4\xi^2=0
\eea
The first three equations are solved by (up to irrelevant sign
ambiguities):
\be
\varphi=(v,v,v)~~~,~~~~~~~v=-\frac{M}{g}~~~.
\ee
The remaining equations are solved, in general, by:
\be
\varphi'=(0,0,0)~~~,~~~~~~~\xi=0~~~,
\ee
unless some further relation is imposed on the coefficients
$g_1,...,g_4$. If $g_2=0$, then, up to an irrelevant reordering, 
we have
\bea
\varphi'&=&(v',0,0)~~~,~~~~~~~~~~~~~~\nn\\
\xi&=&u=-\dd\frac{g_3}{g_4}(\varphi'\varphi')~~~~~~~~~~
\eea
with $v'$ and $u$ undetermined. In this case we find that, for $m^2_{\varphi_0},m^2_{\varphi'_0},
m^2_{\xi_0}>0$, the driving fields $\varphi_0$, $\varphi'_0$
and $\xi_0$ vanish at the minimum. Moreover, if $m^2_{\varphi'},
m^2_\xi<0$, then $u$ and $v'$ slide to large scales, eventually
stabilized by one-loop radiative corrections.
The supersymmetric case is better than the non-supersymmetric
case in two respects. First of all, at least from a technical
viewpoint, the absence of a term in the superpotential is
radiatively stable. Moreover, as we have seen, once $g_2$
has been set to zero, the equations selecting (\ref{align})
as the correct minimum are consistent.

\end{document}